\documentclass[11pt,a4paper]{article}
\usepackage{jcappub}

\pdfoutput=1 

\title{Initial conditions for the scalaron dark matter}

\author{Yuri Shtanov}

\affiliation{Bogolyubov Institute for Theoretical Physics, \\ Metrologichna St.\@ 14-b, Kiev 03143, Ukraine} %
\affiliation{Astronomical Observatory, Taras Shevchenko National University of Kiev, \\ Observatorna St.\@ 3, Kiev 04053, Ukraine} %

\emailAdd{shtanov@bitp.kiev.ua}

\abstract{The scalaron of the metric $f(R)$ gravity can constitute dark matter if its mass is in the range $4\,\text{meV} \lesssim m \lesssim 1\,\text{MeV}$. We give an overview of such $f (R)$ gravity theory minimally coupled to the Standard Model. Similarly to other dark-matter models based on scalar fields, this model has the issue of initial conditions. Firstly, the initial conditions for the scalaron are to be tuned in order to produce the observed amount of dark matter.  Secondly, the primordial spatial inhomogeneities in the field are to be sufficiently small because they generate entropy (or isocurvature) perturbations, which are constrained by observations.  We consider these issues in the present paper. The initial conditions for the scalaron presumably emerge at the inflationary stage. We point out that the homogeneous part of the scalaron initial value is largely unpredictable because of quantum diffusion during inflation. Thus, to account for the observed amount of dark matter, one has to resort to anthropic considerations. Observational constraints on the primordial spatial inhomogeneity of the scalaron are translated into upper bounds on the energy scale of inflation, which happen to be low but not too restrictive.}


\arxivnumber{2207.00267}

\begin{document} 
\maketitle
\flushbottom


\section{Introduction}

It was suggested that dark matter in our universe can be explained in frames of models of $f (R)$ gravity \cite{Capozziello:2006uv, Nojiri:2008nt, Cembranos:2008gj, Cembranos:2015svp, Corda:2011aa, Katsuragawa:2016yir, Katsuragawa:2017wge, Yadav:2018llv, Parbin:2020bpp, Shtanov:2021uif, KumarSharma:2022qdf} (see also reviews \cite{Sotiriou:2008rp, DeFelice:2010aj}). In these models, a new weakly interacting degree of freedom (the scalaron) arises in the gravitational sector, which can be associated with dark matter. 

Among various proposals in this direction, we will focus here on a simple suggestion, pioneered in \cite{Cembranos:2008gj} and recently revisited in our paper \cite{Shtanov:2021uif}, of $f(R)$ models in which essential role is played by the term quadratic in curvature, and which are minimally coupled to the Standard Model of particle physics.  Dark matter in this model is formed of a scalaron field oscillating in a close neighbourhood of the minimum of its potential. The scalaron very weakly interacts with the rest of matter, which ensures its stability.

One of the issues of this model is the issue of initial conditions for the scalaron field. In \cite{Cembranos:2008gj}, it was assumed that the scalaron is initially shifted from the vacuum value in the very early universe with its evolution `frozen' because of the large Hubble friction. The initial value of the scalaron field and its mass are then two free parameters that should be tuned simultaneously in order to fit the observable abundance of dark matter in the late-time universe. The situation here is no different from numerous scalar-field dark-matter models, in which such tuning of initial conditions is always present.  Possible spatial inhomogeneity of the scalaron initial value were not discussed in that paper.  In \cite{Shtanov:2021uif}, we suggested a special initial condition for the scalaron field, which was assumed to reside initially at the minimum of its {\em effective\/} potential that arises due to its interaction with the Higgs field. In this case, there is only one free parameter to be tuned, namely, the scalaron mass, and the scalaron initial value is automatically spatially homogeneous. 

In both scenarios, the spatial homogeneity of the initial value of the scalaron ensures the absence of non-adiabatic (isocurvature) primordial perturbations of dark matter, which are also strongly constrained by observations. However, this assumption of primordial spatial homogeneity is quite idealistic.  In realistic cosmological scenarios, the scalaron is going to be spatially perturbed, so that isocurvature perturbations are going to be present. One, therefore, needs to ensure that the initial spatial perturbations of the scalaron are within the observational bounds on the isocurvature perturbations of dark-matter. The initial conditions for the scalaron are formed presumably at the inflationary era, to which we thus need to resort. Here, one inevitably enters into the realm of unknown physics; however, there are some generic features that can be expected and explored. This will be the subject of the present paper.

After a brief review of the $f(R)$ model of dark matter, we will consider a simple inflationary model based on a scalar field (inflaton) minimally coupled to gravity in the Jordan frame. In the Einstein frame, such an inflaton will strongly modify the potential for the scalaron field during inflation, typically making it monotonic. In this case, the initial conditions for the scalaron will be determined by its random quantum diffusion during inflation and will be largely unpredictable. We will consider then two limiting cases, in which the initial value of the scalaron after inflation happens to be either relatively close to or relatively far from the minimum of its post-inflationary effective potential. The first option is a realistic version of the scenario proposed in \cite{Shtanov:2021uif}, while the second one is close to the scenario proposed in \cite{Cembranos:2008gj}. In both cases, we estimate the amplitude of the isocurvature mode of primordial dark-matter density perturbations and obtain upper bounds on the energy scale of inflation for which this amplitude is within the current observational bounds. These bounds on the inflationary energy scale turn out to be low but not too restrictive, preserving the viability of the dark-matter model under consideration.

\section{$f(R)$ gravity theory}

Working in the metric signature $(-, +, +, +)$, we assume that the Lagrangian of an $f (R)$ gravity theory can be expanded in the form of power series in the scalar curvature $R$:
\begin{equation} \label{Sgs}
L_g = \frac{M^2}{3} f (R) \, , \qquad f (R) = - 2 \Lambda + R + \frac{R^2}{6 m^2} + \ldots \, .
\end{equation} 
Here, $M = \sqrt{3 / 16 \pi G} \approx 3 \times 10^{18}\, \text{GeV}$ is a conveniently normalised Planck mass, and $\Lambda \approx \left( 3 \times 10^{-33}\,\text{eV} \right)^2$ is the cosmological constant in the natural units $\hbar = c = 1$.

The theory with terms up to $R^2$ in \eqref{Sgs} corresponds to the Starobinsky model \cite{Starobinsky:1980te, Vilenkin:1985md}, in which one sets $m \simeq 10^{-5} M$ to account for an inflationary stage with the primordial power spectrum in agreement with current observations \cite{Planck:2018jri}. In this case, the term quadratic in curvature can be regarded as a quantum correction to the effective action for gravity, arising after integrating out of some matter degrees of freedom. In this paper, following \cite{Cembranos:2008gj, Shtanov:2021uif}, we use model \eqref{Sgs} to describe dark matter rather than inflation, hence, we are free to admit other terms in the expansion of \eqref{Sgs} and to allow $m$ to be much smaller than the cited value (in what follows, we will see its value to be confined between meV and MeV by order of magnitude). 

The constant $m$ in this case is regarded as a genuine constant in the action for gravity. The assumed smallness of $m$ (or, in other words, the large dimensionless factor of the $R^2$ term, of the order between $10^{41}$ and $10^{58}$) can be viewed on the same footing as the extreme relative smallness of the cosmological constant $\Lambda$ in the gravitational action, the reason for which is also unknown. What is surprising and somewhat unexpected is that, in spite of such a large factor of the $R^2$ term in the action, the theory well reproduces the Einstein gravity in the domain where it has been tested, giving a candidate for dark matter as an extra bonus (see below). 

Proceeding from the Jordan frame to the Einstein frame, we first write the action with Lagrangian \eqref{Sgs} in the form
\begin{equation}\label{Sg1}
S_g = \frac{M^2}{3} \int d^4 x \sqrt{-g}\, \bigl[ \Omega R - \mu (\Omega) \bigr] \, ,
\end{equation}
where $\Omega$ is a new dimensionless scalar field, and the function $\mu (\Omega)$ is chosen so that variation with respect to $\Omega$ and its substitution into the action returns the original action: 
\begin{align} \label{inv1}
\mu' (\Omega) &= R \quad \Rightarrow \quad \Omega = \Omega (R) \, , \\
f (R) &= \bigl[ \Omega R - \mu (\Omega) \bigr]_{\Omega = \Omega (R)} \, . \label{inv2}
\end{align}
Thus, $f (R)$ is the Legendre transform of $\mu (\Omega)$, with the inverse transform being 
\begin{align} \label{dir1}
f' (R) &= \Omega \quad \Rightarrow \quad R = R (\Omega) \, , \\
\mu (\Omega) &= \bigl[ \Omega R - f (R) \bigr]_{R = R (\Omega)} \, , \label{dir2}
\end{align}
which enables one to find $\mu (\Omega)$ given $f (R)$.

As a final step, we make a conformal transformation in \eqref{Sg1}:
\begin{equation} \label{om}
g_{\mu\nu} \to \Omega^{-1} g_{\mu\nu} \, , \qquad \Omega = e^{\phi / M} \, ,
\end{equation}
where $\phi$ is a new field (the scalaron) parametrising $\Omega$. Action \eqref{Sg1} then becomes that of the Einstein gravity with a scalar field (scalaron) $\phi$. The Lagrangian in the Einstein frame is 
\begin{equation}\label{Sg3}
L_g =  \frac{M^2}{3} R - \frac12 \left( \partial \phi \right)^2 - V (\phi) \, ,
\end{equation}
where the scalaron potential $V (\phi)$ is calculated by using \eqref{Sg1} and \eqref{dir2}:
\begin{equation} \label{V}
V (\phi) = \frac{M^2}{3} e^{- 2 \phi / M} \mu \left( e^{\phi / M} \right) \, .
\end{equation}

Using relations \eqref{dir1} and \eqref{dir2}, it is easy to establish that the scalaron potential has extrema, with $V' (\phi) = 0$, at the Jordan-frame values of $R$ that satisfy
\begin{equation}
R f' (R) = 2 f (R) \, .
\end{equation}
The scalaron mass squared, $m_\phi^2 = V'' (\phi)$, at such an extremum is given by
\begin{equation}
m_\phi^2 = \frac13 \left[ \frac{1}{f''(R)} - \frac{R}{f'(R)} \right] = \frac13 \left[ \frac{1}{f''(R)} - \frac{R^2}{2 f (R)} \right] \, .
\end{equation}
If $m_\phi^2 > 0$, then this is a local minimum. As can be seen from \eqref{dir2}, the scalaron potential varies on a typical scale of the Planck mass $M$.  Hence, in the neighbourhood of the minimum, it is well approximated by a quadratic form on field scales much smaller than $M$.  

For a small cosmological constant, $\Lambda \ll m^2$, the theory has a local minimum at $\phi / M \approx 4 \Lambda / 3 m^2$, corresponding to $R \approx 4 \Lambda$ in the Jordan frame, and the scalaron mass at this minimum is $m_\phi^2 = m^2 + {\cal O} (\Lambda)$. In what follows, we neglect the small cosmological constant in \eqref{Sgs}, which is responsible for dark energy but not for dark matter. In this approximation, the local minimum is situated at $\phi = 0$ (corresponding to $R = 0$ in the Jordan frame), and the scalaron mass at this minimum is $m_\phi = m$. 

The simplest non-trivial model is described by 
\begin{equation}\label{fstar}
f (R) = R + \frac{R^2}{6 m^2} \, .
\end{equation}
In the Einstein frame, it produces Lagrangian \eqref{Sg3} with
\begin{equation} \label{Vstar}
V (\phi) = \frac12  m^2 M^2 \left( 1 - e^{- \phi / M} \right)^2 \, . 
\end{equation}
This potential has an infinitely extended plateau at $\phi \gg M$. 

\begin{figure}[ht]
\begin{center}
\includegraphics[width=.6\textwidth]{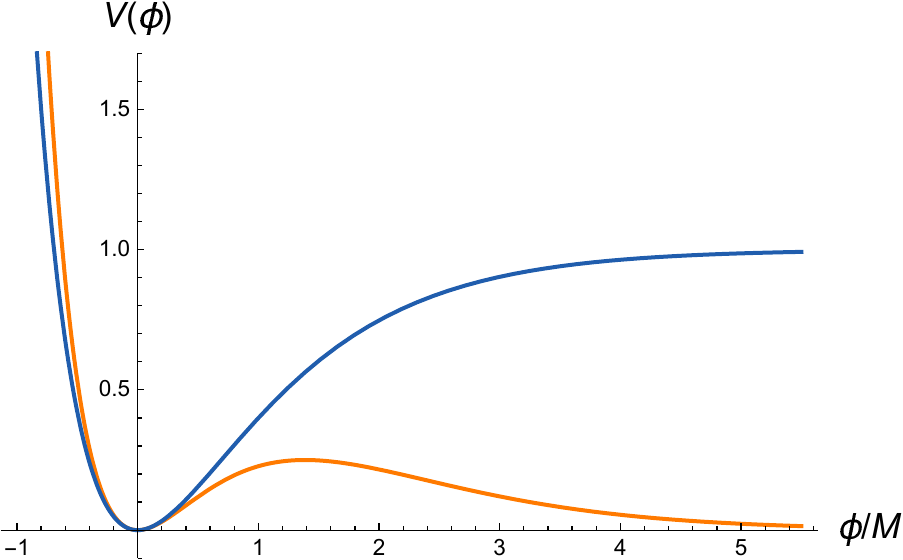}
\caption{Scalaron potentials \eqref{Vstar} (blue) and \eqref{Vrat} (orange) are plotted in units $M^2 m^2/2$. In the region $|\phi|/M \ll 1$, they both are approximated by the quadratic form with mass $m$. \label{fig:pot}}
\end{center}
\end{figure}

Scalaron potentials corresponding to $f (R)$ with higher powers of $R$ in their expansion may have qualitatively different behaviour at large values of $\phi > 0$.  As a simple example, consider
\begin{equation} \label{fexpon}
f (R) = \frac{R}{1 - R / 6 m^2} = R + \frac{R^2}{6 m^2} + \frac{R^3}{36 m^4} + \ldots \, .
\end{equation}
Equation \eqref{dir1} in this case reads
\begin{equation} 
\left(1 - \frac{R}{6 m^2} \right)^{-2} = \Omega \, .
\end{equation}
It has two branches of solutions with respect to $R$. The one which contains the stable critical point $R = 0$ is given by 
\begin{equation}
R (\Omega) = 6 m^2 \left( 1 - \frac{1}{\sqrt{\Omega}} \right) \, .
\end{equation}
The scalaron potential \eqref{V} for this model is then calculated to be
\begin{equation} \label{Vrat}
V (\phi) = 2 M^2 m^2 e^{- \phi / M}  \left( 1 - e^{- \phi / 2 M} \right)^2 \, .
\end{equation}
It has a local maximum at $e^{\phi/M} = 4$ and exponentially decreases as $\phi \to \infty$. Potentials \eqref{Vstar} and \eqref{Vrat} are plotted in figure~\ref{fig:pot}. 

In this paper, we will deal with the potential $V (\phi)$ only in the domain $|\phi| \ll M$, where all such potentials are well approximated by the quadratic form with mass $m$. The initial conditions for the scalaron in the early universe should be limited to the stability region in the neighbourhood of the minimum of its potential.  The initial conditions will be discussed in section~\ref{sec:inflation}.

\section{Coupling to the Standard Model}

As regards the matter part of the action, we assume that it has the usual form of the Standard Model in the Jordan frame. Proceeding to the Einstein frame affects this action as well. Note, however, that most of the Standard Model action is classically conformally invariant (with proper conformal transformation of the matter fields), and, therefore, will retain its original form. The only part that breaks classical conformal invariance is the Higgs sector,\footnote{The neutrino sector extending the Standard Model can also be classically conformally non-invariant, e.g., if it contains Majorana mass terms. We will not go into these details, which are not crucial for this work.} with the Lagrangian
\begin{equation}\label{Sh}
L_h = - \left( D_\mu \Phi \right)^\dagger D^\mu \Phi - \frac{\lambda}{4} \left( 2 \Phi^\dagger \Phi - v^2 \right)^2 \, .
\end{equation}
Here, $D_\mu$ is the gauge covariant derivative involving the SU(2) and U(1) electroweak gauge fields and acting on the Higgs doublet $\Phi$, and
$v \approx 246\, \text{GeV}$ is the symmetry-breaking parameter. After the conformal transformation \eqref{om} accompanied by the Higgs-field transformation $\Phi \to \Omega^{1/2} \Phi$, this Lagrangian becomes\footnote{Similar result of the conformal transformation of fields was under discussion, e.g., in \cite{Burrage:2018dvt}.}
\begin{equation}\label{Shn}
L_h = - \left( D_\mu \Phi \right)^\dagger D^\mu \Phi - \frac{1}{2 M} \partial_\mu \left( \Phi^\dagger \Phi \right) \partial^\mu \phi - \frac{1}{4 M^2} \Phi^\dagger \Phi\, \left( \partial \phi \right)^2 - \frac{\lambda}{4} \left( 2 \Phi^\dagger \Phi - e^{- \phi / M} v^2 \right)^2 \, . 
\end{equation}
We observe the appearance of non-renormalisable interactions of the scalaron $\phi$ with the Higgs field in \eqref{Shn}, which, however, are all suppressed by inverse powers of the large Planck mass $M$.  

The scalaron and the Higgs field are slightly mixed in this model. Choosing the canonical unitary gauge for the Higgs doublet $\Phi$ and shifting it by its vacuum expectation value $v$, we write
\begin{equation}\label{Hshift}
\Phi = \frac{1}{\sqrt{2}} \begin{pmatrix} 0 \\ h \end{pmatrix} \, , \qquad h = v + \varphi \, ,
\end{equation}
where $\varphi$ is the shifted real-valued Higgs field. The quadratic part of the Lagrangian in \eqref{Sg3} and \eqref{Shn} is then
\begin{equation}\label{L2}
L_2 = - \frac12 \left( \partial \varphi \right)^2 - \frac12 \left( 1 + \frac{v^2}{4 M^2} \right) \left( \partial \phi \right)^2 - \frac{v}{2 M} \partial \varphi \partial \phi - \lambda v^2 \left( \varphi + \frac{v}{2 M} \phi \right)^2 - \frac12 m^2 \phi^2 \, .
\end{equation}
The shift of the Higgs field
\begin{equation}\label{shift}
\varphi = \chi - \frac{v}{2 M} \phi 
\end{equation}
brings \eqref{L2} to the diagonal form
\begin{equation}
L_2 = - \frac12 \left( \partial \chi \right)^2 - \frac12 \left( \partial \phi \right)^2 - \lambda v^2 \chi^2 - \frac12 m^2 \phi^2 \, .
\end{equation}
The Higgs field has mass $m_\text{H} = \sqrt{2 \lambda} v \approx 125\,\text{GeV}$ in the Standard Model, so that $\lambda \approx 0.13$.

As was mentioned above, interactions between the Higgs field and the scalaron in \eqref{Shn} are suppressed by inverse powers of the Planck mass $M$. Thus, for instance, the lowest-order interaction terms that allow for the Higgs boson $\chi$ to decay into a pair of scalarons have the form
\begin{equation}
L_\text{int} = \frac{v}{4 M^2} \phi\, \partial_\mu \chi\, \partial^\mu \phi - \frac{\lambda v^3}{4 M^2} \chi \phi^2 \, .
\end{equation}
They will make a negligible contribution $\Gamma_{\chi \to \phi \phi} \sim 10^{-62}$~MeV to the total decay width $\Gamma_\chi \approx 4$~MeV \cite{Higgs:2016ypw} of the Higgs boson. This is another manifestation of the fact that the scalaron is a gravitational degree of freedom, and creation of its quanta is suppressed by inverse powers of $M$ similarly to creation of gravitons.

Shift \eqref{shift} generates couplings between the scalaron field $\phi$ and matter fields of the Standard Model.  Thus, the usual Yukawa coupling $- \gamma h \bar \psi \psi$ between the original Higgs field $h$ and a fermion $\psi$, where $\gamma$ is the Yukawa coupling constant, in view of \eqref{Hshift} and \eqref{shift} will generate the coupling
\begin{equation}\label{psicoup}
\frac{\gamma v}{2 M} \phi \bar \psi \psi = \frac{m_\psi}{2 M} \phi \bar \psi \psi \, ,
\end{equation}
where $m_\psi = \gamma v$ is the fermion mass. Likewise, the coupling of the Higgs field and vector gauge fields in \eqref{Sh} will generate the couplings
\begin{equation}\label{gcoup}
- \frac{m_W^2}{M} \phi W_\mu^+ W^{-\mu} - \frac{m_Z^2}{2 M} \phi Z_\mu Z^\mu
\end{equation} 
of the scalaron to the vector bosons $W^\pm$ and $Z$. These couplings would allow for the scalaron decays into other particles provided its mass $m$ is sufficiently large. In particular, interactions \eqref{psicoup} can allow for the scalaron decays into electron-positron pairs if $m > 2 m_e$. For the scalaron representing all of dark matter in the universe, an upper bound on its mass was derived from the observed 511~keV line emission from the Galactic Centre \cite{Cembranos:2008gj, Cembranos:2015svp}:
\begin{equation}\label{mup}
m \lesssim 1.2\, \text{MeV} \, .
\end{equation}

Thus far, we were discussing local conformal transformations at the classical level; however, the Standard Model as a quantum field theory breaks local conformal symmetry also at the quantum level. Conformal transformation \eqref{om} and the corresponding matter-field redefinition affect the implicit dimensionful parameters such as the $\Lambda_\text{QCD}$ scale. Together with the scalaron-dependence of the vacuum expectation value of the Higgs field in \eqref{Shn}, this leads to the scalaron-dependence of all particle masses in the Einstein frame, making the original Jordan-frame metric the observable one, in accordance with the equivalence principle.  These issues are discussed in detail in our paper \cite{Shtanov:2022wpr} to which we refer the reader.

Quantum breaking of the local conformal symmetry results in additional interactions between the scalaron and gauge bosons due to the conformal (or scale) anomaly \cite{Cembranos:2008gj, Katsuragawa:2016yir}. Their origin can be traced to the special Jacobian arising in the transformation of the path-integral measure of fermionic fields from the Jordan to the Einstein frame \cite{Katsuragawa:2016yir}. These interactions have the form
\begin{equation}\label{anom}
\left( C \alpha_\text{g} \frac{\phi}{M} + \ldots \right) \text{tr}\, F_{\mu\nu} F^{\mu\nu} \, ,
\end{equation} 
where $\alpha_\text{g} = e_\text{g}^2 / 4 \pi$ is the corresponding gauge coupling constant, $C$ is a numerical constant depending on the number of fermionic fields coupled to gauge bosons, and lower dots denote terms of higher order in $\phi / M$. 

Interactions of the sort \eqref{psicoup}, \eqref{gcoup}, \eqref{anom} allow for the scalaron decays into photons. The lifetime with respect to such decays is estimated as \cite{Katsuragawa:2016yir}
\begin{equation}
\tau \sim \frac{10^2 M^2}{\alpha_\text{em}^2 m^3} \sim 10^{38} \left( \frac{\text{eV}}{m} \right)^3 \, \text{yr} \, , 
\end{equation}
where $\alpha_\text{em} \approx 1/137$ is the electromagnetic coupling constant. This can be compared with the age of the universe $1.4 \times 10^{10}$~yr. For the upper bound \eqref{mup} on the scalaron mass, such decays would be difficult to detect in the isotropic diffuse photon background \cite{Cembranos:2008gj, Cembranos:2015svp}.

The minimal coupling of the metric to matter in the Jordan frame results in a universal coupling of the scalaron to matter fields, collectively denoted as $\Psi$, with the Lagrangian density
\begin{equation}
{\cal L}_\text{m} \left( e^{- \phi / M} g_{\mu\nu}, \Psi \right) \, .
\end{equation}
Here, the fields $\Psi$ are those of the Jordan frame, while the metric is transformed to the Einstein frame. If one takes the Jordan frame as the starting frame in which all matter fields are quantised, then it is the Jordan-frame metric $e^{- \phi / M} g_{\mu\nu}$ that plays the role of the `observable' metric in the Einstein frame \cite{Dicke:1961gz, Faraoni:2006fx, Shtanov:2022wpr}, and an exchange of the scalaron $\phi$ then produces an additional universal gravitational force of Yukawa type, with the total gravitational potential \cite{Stelle:1977ry}
\begin{equation}
\Phi_\text{grav} = - \frac{2 G {\cal M}}{r} \left( 1 + \frac13 e^{- m r} \right) \, .
\end{equation}
Non-observation \cite{Kapner:2006si, Adelberger:2006dh} of such additional Yukawa forces between non-relativistic masses at small distances leads to a lower bound on the scalaron mass (see also \cite{Cembranos:2008gj, Cembranos:2015svp, Perivolaropoulos:2019vkb})
\begin{equation}\label{mlow}
m \geq 2.7 \times 10^{-3}\, \text{eV} \quad \text{at 95\% C.L.}
\end{equation}

Finally in this section, we briefly discuss the issue of quantum corrections to the value of the scalaron mass $m$. Interaction of the Yukawa type \eqref{psicoup} results in a finite (after minimal subtraction of divergence) one-loop contribution to the scalaron self-energy of the order
\begin{equation}
\Pi \simeq \left( \frac{m_\psi}{2 \pi M} \right)^2 m_\psi^2
\end{equation}
at small momentum invariant $k^2$.  Among the fermions of the Standard Model, the heaviest is the top quark with mass $m_\psi \approx 172$~GeV, and one expects quantum corrections to the scalaron squared mass of the order of $\left( 10^{- 6}\, \text{eV} \right)^2$. This is negligible given the lower bound \eqref{mlow}. Analysis of other interactions leads to the same conclusion.  Thus, the scalaron mass is free from the naturalness issue within the Standard Model because of the extremely weak scalaron couplings. This is no longer the case if there exist much heavier particles (such as an inflaton, which we consider in the next section) beyond the Standard Model with the same universal coupling to the scalaron.\footnote{The naturalness issue in quantum field theory \cite{Dine:2015xga}, perhaps, is not that problematic as usually deemed, being dependent on the renormalisation scheme. Other approaches to renormalisation may be free from it \cite{Epstein:1973gw, Scharf:2014, Mooij:2021ojy, Mooij:2021lbc}.}

\section{Inflation and initial conditions for the scalaron}
\label{sec:inflation}

Initial conditions for the scalaron are supposed to be formed during the inflationary epoch. One can envisage many different models of inflation in which our model of dark matter can be embedded. First of all, one might try to avoid introducing new fields responsible for inflation but instead to arrange the coupling $\propto h^2 R$ of the Higgs field to the scalar curvature in the Jordan frame. In this case, one obtains the so-called Higgs--scalaron inflation \cite{Salvio:2015kka, Wang:2017fuy, Ema:2017rqn, He:2018gyf, Gorbunov:2018llf, Gundhi:2018wyz, Canko:2019mud, Cheong:2021vdb}. However, in models of this kind, either inflation proceeds along the valley of the resulting potential in the scalaron direction, and our scalaron mass $m$ is way too small to sustain viable inflation, or it proceeds in the Higgs-field direction, in which case it is incompatible with observations because of the large Higgs self-coupling constant $\lambda \approx 0.13$. Therefore, in our case, one should resort to an unknown inflationary sector.

Whatever mechanism acted during inflation and subsequent reheating, it is not supposed to produce any appreciable background of gravitational radiation on wavelengths typical to those of the CMB (which would be in conflict with the current theory of nucleosynthesis).  We, therefore, can also reasonably suggest that it did not produce any appreciable background of the scalaron particles with similar de~Broglie wavelengths. However, along with the generation of density and gravitational perturbations outside the Hubble radius during inflation, similar fluctuations of the scalaron field will inevitably be produced, forming an inhomogeneous background with a characteristic power spectrum on large spatial scales. Since the scalaron quantum fluctuations are not correlated with those of other fields, they will constitute the isocurvature perturbations of dark matter, whose magnitude is known to be bounded from above by observations. This bound is to be satisfied in the present scenario.

Typically, inflation is based on a special scalar field (inflaton) $\zeta$ that is arranged to interact with the matter fields in order to ensure post-inflationary reheating. Suppose that its free part is described by the following Lagrangian with minimal coupling to the metric in the Jordan frame:
\begin{equation}
L_\text{infl} = - \frac12 \left( \partial \zeta \right)^2 - W (\zeta) \, ,
\end{equation}
where $W (\zeta)$ is the potential term.  In the Einstein frame, after the conformal transformation $\zeta \to \Omega^{1/2} \zeta$, the Lagrangian takes the form similar to \eqref{Lhs}:
\begin{equation} \label{Linf}
L_\text{infl} = - \frac12 \left( \partial \zeta \right)^2 - \frac{1}{2 M} \zeta \left( \partial \zeta \partial \phi \right) - \frac{1}{8 M^2} \zeta^2\, \left( \partial \phi \right)^2 - e^{- 2 \phi / M} W \left( e^{\phi / 2 M} \zeta \right) \, ,
\end{equation}
where we have also taken into account the transformation \eqref{om} of the metric and the presence of the factor $\sqrt{ - g}$ in the Lagrangian density.

Under the condition $\zeta^2 \lesssim M^2$, the scalaron kinetic term will not be strongly modified, but its potential will receive corrections in view of the last term in \eqref{Linf}.  For instance, if there is a term $\propto \zeta^2$ present in $W (\zeta)$, then it generates the term $\propto e^{- \phi / M} \zeta^2$ in the scalaron potential.  In this case, for large values of $\zeta$ during inflation, the scalaron potential may even have no minimum, monotonically exponentially decreasing as $\phi \to \infty$.\footnote{One should note that the Higgs field also exhibits quantum fluctuations during inflation, of order $\Phi^\dagger \Phi \sim H^2$, which will also modify the scalaron potential in \eqref{Shn}. This effect is typically much smaller than that from the inflaton field.}  The value of the scalaron field in the observable part of the universe will then be a result of random quantum diffusion during inflation \cite{Vilenkin:1983xq, Starobinsky:1986fx}. 

As an example, consider the simplest quadratic potential of the inflaton $ 
W ( \zeta ) = m_\zeta^2 \zeta^2 / 2$. The potential in \eqref{Linf} is then 
\begin{equation} \label{Lalp}
e^{- 2 \phi / M} W \left( e^{\phi / 2 M} \zeta \right) = \frac12 m_\zeta^2 e^{- \phi / M} \zeta^2 \, .
\end{equation}
This potential has a ground-state valley at $\zeta =0$ with the inflaton mass $m_\zeta (\phi) = m_\zeta e^{- \phi / 2 M}$.  Inflation in such a model is based on two scalar fields $\zeta$ and $\phi$, ending up in this valley at an unpredictable value of the scalaron. 

During preheating, the local scalaron mass is $m_\phi (\phi) \simeq m_\zeta e^{- \phi / 2 M} \left\langle \zeta^2 \right\rangle^{1/2} / M$, where the time average over the inflaton oscillations is indicated by brackets.  Since the inflaton oscillates with frequency $m_\zeta (\phi) = m_\zeta e^{- \phi / 2 M}$, it can create the scalaron quanta during preheating when $\left\langle \zeta^2 \right\rangle \ll M^2$. The number density of the scalaron quanta created in this process will depend on the rate of creation of other particles by the inflaton. All these details are strongly dependent on the inflationary model with its interactions, and we assume the number density of the scalaron quanta to be small so that, after preheating, the scalaron energy density is dominated by the classical scalaron field. 

The above example shows that the scalaron field in an inflationary model can end up at any reasonable value of $\phi$.  We will see in section~\ref{sec:gen} that production of the observed abundance of dark matter under constraints \eqref{mup} and \eqref{mlow} requires the condition $|\phi|/M \lesssim 10^{-6}$.  Thus, merely for anthropic reasoning, inflation should end with the scalaron value close to the minimum of its potential in the observable patch of the universe.

By the end of inflation, the classical field $\phi$ is subject to quantum fluctuations on super-Hubble scales, with typically almost scale-invariant power spectrum and with variance (see, e.g., \cite{Baumann-2011})
\begin{equation}
\left\langle \delta \phi^2 \right\rangle \simeq \left(\frac{H_\text{inf}}{2 \pi } \right)^2 \, ,
\end{equation}
where $H_\text{inf}$ is the characteristic Hubble parameter during the final stages of inflation. To ensure the adiabatic character of the primordial power spectrum of dark matter, these fluctuations should be sufficiently small compared to the background value of the scalaron, so that, after dark matter is formed, we should have \cite{Planck:2018jri}
\begin{equation}\label{adcon}
{\cal P}_\phi \equiv \left\langle \left( \frac{\delta \rho_\phi }{\rho_\phi} \right)^2 \right\rangle \lesssim \ 0.02\, {\cal P}_{\cal R} \sim 10^{-11} \, ,
\end{equation}
where ${\cal P}_{\cal R} \sim 10^{-9}$ is the scale-invariant power spectrum of adiabatic perturbations.  This inequality leads to constraints on the inflationary energy density which will be obtained in section~\ref{sec:gen}.

\section{Post-inflationary evolution of the scalaron}

The full Higgs--scalaron Lagrangian in the unitary gauge in the Einstein frame becomes [see \eqref{Sg3} and \eqref{Shn}]
\begin{align}\label{Lhs}
L_\text{HS} = - \frac12 \left( \partial \phi \right)^2 - \frac12 \left( \partial h \right)^2 - \frac{1}{2 M} h\, \left( \partial h \partial \phi \right) - \frac{1}{8 M^2} h^2 \left( \partial \phi \right)^2 \nonumber \\ - V (\phi) - \frac{\lambda}{4} \left( h^2 - e^{- \phi / M} v^2 \right)^2 \, .
\end{align}
In thermal equilibrium of the hot universe, the Higgs field acquires thermal corrections to its potential, which, for large temperatures $T \gtrsim m_i$, where $m_i$ are particle masses caused by interaction with the Higgs field, can be estimated as (see \cite{GR})
\begin{equation}\label{VT}
V_T ( h ) \simeq \frac16 T^2 h^2 - \frac{1}{100} T h^3 \, .
\end{equation}
The numerical coefficients in this expression are not exact; they were chosen rational, and close to the numbers one obtains in the Standard Model, just to simplify the subsequent expression for the critical temperature. Taking into account \eqref{Lhs}, we obtain the effective Lagrangian for the system of the Higgs and scalaron fields:
\begin{align}\label{Leff}
L_\text{eff} &= - \frac12 \left( \partial \phi \right)^2 - \frac12 \left( \partial h \right)^2 - \frac{1}{2 M} h\, \left( \partial h \partial \phi \right) - \frac{1}{8 M^2} h^2 \left( \partial \phi \right)^2 \nonumber \\ &\quad {} - V (\phi) - \frac{\lambda}{4} \left( h^2 - e^{- \phi / M} v^2 \right)^2 - \frac16 T^2 h^2 + \frac{1}{100} T h^3 \, .
\end{align}
Here we have neglected the vacuum quantum corrections to the Higgs effective potential.

The Higgs field strongly interacts with the primordial plasma, and, being in thermal equilibrium, has its expectation value $h$ at the minimum of its effective potential in \eqref{Leff}. The cosmological equation for the spatially homogeneous scalaron is then
\begin{align} \label{eqs}
\ddot \phi + 3 H \dot \phi + \frac{1}{4 M} \left[ \left( h^2 \right)^{\cdot\cdot} + 3 H \left( h^2 \right)^{\cdot} \right] + \frac{1}{4 M^2} \left[ \left( h^2 \dot \phi \right)^\cdot + 3 H h^2 \dot \phi \right] \nonumber \\ {} + V ' (\phi) + \frac{\lambda v^2}{2 M} e^{- \phi / M} \left( h^2 - e^{- \phi / M} v^2 \right) = 0 \, . 
\end{align}
Here, $H = \dot a /a$ is the Hubble parameter, with $a$ being the scale factor in the space-time metric of the expanding universe. 

Before the electroweak crossover, the Higgs-field expectation value is $h = 0$, and the scalaron total potential is
\begin{equation}
V_{0} (\phi) = V (\phi) + \frac{\lambda}{4} v^4 e^{- 2 \phi/M} \, .
\end{equation}
Its minimum $\phi_0$ is thus displaced from the zero value in the positive direction. For potential \eqref{Vstar}, it is given by
\begin{equation}\label{phi0}
\phi_0 = M \ln \left( 1 + \frac{\lambda v^4}{2 m^2 M^2} \right) \approx \frac{\lambda v^4}{2 m^2 M} \ll M \, , 
\end{equation} 
where we have taken into account the inequality which follows from the lower bound \eqref{mlow}:
\begin{equation} \label{ineq2}
\frac{\phi_0}{M} \approx \frac{\lambda v^4}{2 m^2 M^2} \ \lesssim \ 3.6 \times 10^{-6} \ll 1 \, .
\end{equation}
The scalaron effective mass is modified very slightly: 
\begin{equation} \label{mass}
V'' ( \phi_0 ) = \frac{m^2}{1 + \lambda v^4 / 2 m^2 M^2} \approx m^2 \, .
\end{equation}
The approximations in \eqref{phi0} and \eqref{mass} are valid for the generic $f (R)$ model.  

The temperature $T_c$ of the beginning of the electroweak crossover is determined by vanishing of the second derivative of the effective potential in \eqref{Leff} with respect to $h$ at $h = 0$. Performing a simple calculation, one obtains
\begin{equation}\label{Tc}
T_c = \sqrt{3 \lambda} v e^{- \phi / 2 M}  \approx \sqrt{3 \lambda} v \approx 154\, \text{GeV} \, ,
\end{equation} 
where we have used the condition $|\phi| / M \ll 1$.  After the beginning of the electroweak crossover, the Higgs field starts relaxation to the minimum of its effective potential, and its expectation value at the initial stage of this process is given by [we neglect the last term in \eqref{Leff}, which is legitimate at the early stages of crossover]
\begin{equation} \label{relax}
h^2 = v^2 e^{- \phi  / M} - \frac{T^2}{3 \lambda} = v^2  e^{- \phi  / M} \left( 1 - \frac{T^2}{T_c^2} \right) \approx v^2 \left( 1 - \frac{T^2}{T_c^2} \right) \, , \qquad T \leq T_c \, .
\end{equation} 

The scalaron evolves according to equation \eqref{eqs} with $h^2$ equal to zero before the electroweak crossover, and given by \eqref{relax} after it. Such an approximation to the field dynamics can be justified by the fact that the relaxation time of the Higgs field is much smaller than the scalaron time scale $m^{-1}$. Indeed, the Higgs effective mass $m_\text{H} (h) = \sqrt{2 \lambda} h$ during the crossover becomes much larger than $m$ for $h / v \gg m / \sqrt{2\lambda} v \sim  m / 10^{2}\, \text{GeV}$, i.e., practically at the very beginning of crossover for small scalaron masses $m \ll 10^{2}\, \text{GeV}$.

Several terms in equation \eqref{eqs} can be neglected. Thus, when compared to the scalaron mass term $V' (\phi) \approx m^2 \phi$, all terms with time derivatives in \eqref{eqs} containing the Higgs field can be dropped because of the presence of a very small parameter ${v^2}/{M^2} \sim 10^{-32}$. Therefore, we can use the following simple equation for the scalaron:
\begin{equation} \label{eqsf}
\ddot \phi + 3 H \dot \phi + m^2 \phi + \frac{\lambda v^2}{2 M} \left( h^2 - v^2 \right) = 0 \, , 
\end{equation}
in deriving which, we have neglected terms of higher order in $|\phi| / M \ll 1$.

The scalaron is initially frozen until the friction coefficient $3 H$ in \eqref{eqsf} becomes comparable to its mass $m$. After that, it starts oscillating adiabatically, with adiabaticity interrupted only at the beginning of electroweak crossover. The value of the Hubble parameter at the electroweak crossover is
\begin{equation}\label{Hc}
H_c \equiv \left( \frac{\rho_c}{2 M^2} \right)^{1/2} \equiv \left( \frac{\pi^2 g_c T_c^4}{60 M^2} \right)^{1/2} \simeq \left( \frac{3 \pi^2 g_c}{20} \right)^{1/2} \frac{\lambda v^2}{M} \simeq 3 \times 10^{-5}\, \text{eV} \, .
\end{equation}
Here, $g_c \approx 100$ is the number of relativistic degrees of freedom  in thermal equilibrium at this moment. By virtue of \eqref{mlow}, $H_c \ll m$, hence, the Hubble friction term in \eqref{eqsf} is already very small compared to the scalaron mass term at the electroweak crossover.  This means that the scalaron, if it is not initially in the minimum \eqref{phi0} of the potential (this option is also considered below), starts oscillating prior to the electroweak crossover. 

The minimum of the scalaron potential in \eqref{Leff} is reached at
\begin{equation} \label{phi0h}
\phi_0 (h) \approx \phi_0 \left( 1 - \frac{h^2}{v^2} \right) \, .
\end{equation}
It is equal to $\phi_0$ before the electroweak crossover. After the beginning of the crossover, it decreases from unity to the zero value asymptotically. During the initial stage of the electroweak crossover, according to \eqref{relax}, it evolves as $\phi_0 (h) = \phi_0 T^2/T_c^2$. Later on, it deviates from this law because of the deviation of the Higgs thermal potential from the high-temperature approximation \eqref{VT}. In particular, the leading contributions to the effective potential for the Higgs field during crossover contain exponential factors of the form $e^{- m_i / T}$, where $m_i$ are the (temperature-dependent) particle masses \cite{Laine:2016hma}. The expectation value of the Higgs field at temperatures $T \sim m_i$ will decrease with the rate of change 
\begin{equation}
t_h^{-1} \sim \frac{|\dot T| m_i}{T^2} \simeq \frac{H m_i}{T} \, .
\end{equation}
This inverse time scale is of the order $H_c$ at the beginning of crossover, decreasing as $t_h^{-1} \propto T$ with temperature. Since $m t_h \sim m / H_c \gg 1$ initially at the moment of crossover, it remains to be large all the time. Since we also have $H \ll m$, the evolution of the quantity $\phi_0 (h)$ is adiabatic with respect to the mass $m$ all the time after the beginning of crossover. 

Evolution \eqref{eqsf} of the scalaron is characterised by the adiabatic invariant
\begin{equation}\label{I}
I = \frac{a^3}{2} \left[ \dot \xi^2 + m^2 \xi^2 \right] \approx \text{const} \, ,
\end{equation}
where $\xi \equiv \phi - \phi_0 (h)$.  This invariant, representing the comoving energy density of the scalaron oscillations, remains constant as the quantity $\phi_0 (h)$ changes slowly compared to the inverse mass scale $m^{-1}$.  This adiabatic regime breaks only at the beginning of the electroweak crossover, where the quantity $\phi_0 (h)$ acquires a quick (non-adiabatic) push:
\begin{equation}\label{push}
\dot \phi_0 \bigr|_{T = T_c} = - 2 H_c \phi_0 \, .
\end{equation}
This results in an abrupt change of the value of the adiabatic invariant at this moment.  As the quantity $\phi_0 (h)$ eventually tends to zero, the adiabatic invariant \eqref{I} represents the comoving energy density of the scalaron, with physical energy density being $\rho_\phi = I / a^3$.

We denote various quantities at the moment of the start of the scalaron oscillations (where $3 H \simeq m$) by the index `i.' The initial adiabatic invariant \eqref{I} is then
\begin{equation}\label{Ini}
I_\text{i} = \frac{a_\text{i}^3}{2} m^2 \left( \phi_\text{i} - \phi_0 \right)^2 \, .
\end{equation}

\subsection{Special initial condition}
\label{sec:spec}

The scenario where the scalaron initially is at rest at the minimum of its effective potential ($\phi_\text{i} = \phi_0$)  was considered in \cite{Shtanov:2021uif}.  In this case, initially we have $I_\text{i} = 0$, and the adiabatic invariant after the electroweak crossover is determined from \eqref{I} and \eqref{push} as
\begin{equation}
I = 2 a_c^3 H_c^2 \phi_0^2 \, .
\end{equation}
It preserves its value at later times, so that, today, the energy density of the scalaron oscillations is given by
\begin{equation}
\rho_\phi = 2 \phi_0^2 H_c^2 \left( \frac{a_c}{a} \right)^3 = \frac{2 \pi^2 \phi_0^2 g_c T_c^4}{15 M^2} \left( \frac{a_c}{a} \right)^3 \, .
\end{equation}

For the present epoch with the scale factor $a_0$, we have $ \left( {a_c}/{a_0} \right)^3 = {2 T_0^3}/{g_c T_c^3}$, where $T_0 = 2.34 \times 10^{-4}\, \text{eV}$ is the temperature of the cosmic microwave background. The present energy density of the scalaron is then
\begin{equation}
\rho_0 = \frac{4 \pi^2 \phi_0^2 T_c T_0^3}{15 M^2} \simeq \frac{\pi^2 \lambda^{5/2} v^9 T_0^3}{5 \sqrt{3} m^4 M^4} \,  ,
\end{equation}
and the corresponding cosmological parameter
\begin{equation}
\Omega_\phi = \frac{\rho_0}{2 M^2 H_0^2} \simeq \frac{\pi^2 \lambda^{5/2} v^9 T_0^3}{10 \sqrt{3} m^4 M^6 H_0^2} \,.
\end{equation}
Substituting here the values of the physical quantities, we obtain
\begin{equation} \label{Om}
\Omega_\phi h_{100}^2 \simeq 0.12 \left( \frac{4.4 \times 10^{-3} \, \text{eV}}{m} \right)^4 \, ,
\end{equation}
where $h_{100} = H_0 / 100\, \text{km}\,\text{s}^{-1}\text{Mpc}^{-1}$. Thus, for 
\begin{equation} \label{m}
m \approx 4.4 \times 10^{-3} \, \text{eV} \, ,
\end{equation}
we obtain the result consistent with the established \cite{Aghanim:2018eyx} abundance of dark matter, $\Omega_c h_{100}^2 = 0.12$.  For this scenario, the value of $m$ is completely fixed, and $\phi_0 \approx 4.8 \times 10^{12}$~GeV according to \eqref{phi0}. 

\subsection{Generic initial condition}
\label{sec:gen}

In the scenario considered in \cite{Shtanov:2021uif} and described in section~\ref{sec:spec}, the scalaron initially resides at the minimum of its effective potential.  However, it is not clear by which mechanism such an initial condition might be realised, given that the scalaron interacts weakly with the rest of matter and has a relatively small mass.  In the inflationary model, discussed in section~\ref{sec:inflation}, the scalaron is subject to quantum fluctuations and, in general, is displaced from the minimum of its effective potential after inflation, so that the initial value \eqref{Ini} of the adiabatic invariant is nonzero.  In this case, the scalaron unfreezes and starts oscillating as soon as the value of the decreasing friction coefficient $3 H$ in \eqref{eqsf} reaches roughly the value of $m$.  For the early hot universe, this condition reads
\begin{equation}\label{Hm}
3 H_\text{i} \equiv \sqrt{\frac{3 g_\text{i}}{5}} \frac{\pi T_\text{i}^2}{M} \simeq m \, ,
\end{equation}
where $g_\text{i} \approx 100$ is the number of relativistic degrees of freedom in thermal equilibrium at this moment.  With the lower bound \eqref{mlow} on $m$, this gives $T_\text{i} \gtrsim g_\text{i}^{-1/4}$~TeV, so that oscillations start well before the electroweak crossover.

Let $\xi_-$ and $\xi_+$ be the amplitudes of oscillations of $\xi \equiv \phi - \phi_0$ just before and right after the beginning of the electroweak crossover, respectively. The adiabatic invariant \eqref{I} experiences a jump at this moment because of the velocity push \eqref{push}. A typical velocity of the scalaron at this moment is $\dot \phi \sim m \xi_-$.  The relative change in the adiabatic invariant will depend on the relation between this value and \eqref{push}.  

Under a special situation where the velocity push \eqref{push} exactly matches the velocity of the scalaron, an occasional cancellation of velocities during the electroweak transition is possible, reducing the adiabatic invariant to zero. Disregarding this special and improbable situation, there are two limiting cases to be considered. 

{\bf 1.}~The adiabatic invariant greatly increases at the electroweak crossover if 
\begin{equation}\label{ximc}
m \xi_- \ll 2 H_c \phi_0 \approx m \xi_+ \, .
\end{equation}
In this case, we come close to the scenario of section~\ref{sec:spec}, in which the scalaron is stationary before the electroweak crossover. Let us see under what conditions this is realised. 

The amplitude $\xi_-$ just before the electroweak crossover is related to the amplitude $\xi_\text{i}$ at the moment of the beginning of oscillations as
\begin{equation}\label{xirel}
\xi_- = \xi_\text{i} \left( \frac{a_\text{i}}{a_c} \right)^{3/2} = \xi_\text{i} \left( \frac{H_c}{H_\text{i}} \right)^{3/4} \simeq \xi_\text{i} \left( \frac{3 H_c}{m} \right)^{3/4} \, .
\end{equation}
Inequality \eqref{ximc} then translates into
\begin{equation}
\xi_\text{i} \ll \left( \frac{H_c}{m} \right)^{1/4} \phi_0 \lesssim 3 \phi_0 \, ,
\end{equation}
where we have used \eqref{mlow} and \eqref{Hc}.  Thus, in order that the scenario of section~\ref{sec:spec} be realised, it suffices that the initial amplitude $\xi_\text{i} = |\phi_\text{i} - \phi_0|$ be much smaller than the minimum value $\phi_0$ given by \eqref{phi0}.

The adiabaticity constraint \eqref{adcon} on the inflationary Hubble parameter $H_\text{inf}$ can be taken into account immediately after the beginning of electroweak crossover.  It then reads
\begin{equation}\label{xicon}
\left\langle \left( \frac{ \delta I_+ }{I_+} \right)^2 \right\rangle \lesssim 10^{-11} \, ,
\end{equation}
where $I_+$ is the adiabatic invariant after the electroweak crossover, and $\delta I_+$ is its quantum fluctuation on super-Hubble spatial scales.  We have
\begin{align}
I_+ &= \frac{a_c^3}{2} m^2 \xi_+^2 \approx 2 a_c^3 H_c^2 \phi_0^2 \, , \\  \delta I_+  &= \frac{a_c^3}{2} \left[ \left( m \xi_+ + m \delta \xi_- \right)^2 - m^2 \xi_+^2 \right] \approx a_c^3 m^2 \xi_+ \delta \xi_- \approx 2 a_c^3 m H_c \phi_0 \delta \xi_-\, .
\end{align}
The fluctuations $\delta \xi_-$ and $\delta \xi_\text{i}$ are related similarly to \eqref{xirel}.  Since we also have $\left\langle \delta \xi_\text{i}^2 \right\rangle = \left( H_\text{inf} / 2 \pi \right)^2$, from \eqref{xicon} we obtain the constraint
\begin{equation}\label{H1}
\left( \frac{H_\text{inf}}{2 \pi  \phi_0} \right)^2 \left( \frac{27 m}{H_c} \right)^{1/2} \lesssim 10^{-11} \quad \Rightarrow \quad H_\text{inf} \lesssim 10^7~\text{GeV} \, ,
\end{equation}
where, in the last estimate, we have used \eqref{phi0} and \eqref{m}. Thus, inflationary origin of perturbations in this case require inflation with the reheating temperature
\begin{equation}\label{Tr1}
T_r \lesssim 10^{12}~\text{GeV} \, .
\end{equation}

{\bf 2.}~The opposite limiting case is where the adiabatic invariant changes insignificantly during the electroweak crossover, which is realised when $m \xi_- \simeq m \xi_+ \gg 2 H_c \phi_0$, hence
\begin{equation}\label{xil}
\xi_\text{i} \gg \left( \frac{H_c}{m} \right)^{1/4} \phi_0 \, .
\end{equation}
Using the conservation of the adiabatic invariant \eqref{I}, in this case, we obtain
\begin{equation} \label{r0}
\rho_0 = \rho_\text{i} \left( \frac{a_\text{i}}{a_0} \right)^3 = \frac{m^2 \xi_\text{i}^2}{2} \cdot \frac{2 T_0^3}{g_\text{i} T_\text{i}^3} = \left( \frac{ 3 \pi^2}{5} \right)^{3/4} \frac{m^{1/2} \xi_\text{i}^2 T_0^3}{g_\text{i}^{1/4} M^{3/2}} \, ,
\end{equation}
where we have used \eqref{Hm} to express $T_\text{i}$ through $m$.  The result \eqref{r0} involves two free parameters, the mass $m$ and the initial amplitude $\xi_\text{i}$, which are to be tuned in order to obtain the observable dark-matter abundance.  The cosmological parameter is
\begin{equation}
\Omega_\phi = \frac{\rho_0}{2 M^2 H_0^2} = \frac12 \left( \frac{ 3 \pi^2}{5} \right)^{3/4} \frac{m^{1/2} \xi_\text{i}^2 T_0^3}{g_\text{i}^{1/4} M^{7/2} H_0^2} \, .
\end{equation}
Then
\begin{equation}
\Omega_\phi h_{100}^2 = 0.12 \left( \frac{100}{g_\text{i}} \right)^{1/4} \left( \frac{m}{\text{eV}} \right)^{1/2} \left( \frac{\xi_\text{i}}{10^{-7}\, M} \right)^2 \, .
\end{equation}
Since, observationally, $\Omega_c h_{100}^2 = 0.12$ \cite{Aghanim:2018eyx}, one obtains the constraint in the $\{ m, \xi_\text{i}\}$ parametric space:
\begin{equation}\label{con}
\left( \frac{100}{g_\text{i}} \right)^{1/4} \left( \frac{m}{\text{eV}} \right)^{1/2} \left( \frac{\xi_\text{i}}{10^{-7}\, M} \right)^2 = 1 \, .
\end{equation}
Condition \eqref{xil} together with \eqref{phi0} and \eqref{con} implies
\begin{equation} \label{mll}
m \gg 4 \times 10^{-3} \left( \frac{100}{g_\text{i}} \right)^{1/16} \, \text{eV} \, .
\end{equation}

In view of the upper bound \eqref{mup} and the lower bound \eqref{mll} on the mass $m$, from \eqref{con} we have the constraint (setting $g_\text{i} = 100$)
\begin{equation}
3 \times 10^{-9} \lesssim \frac{\xi_\text{i}}{M} \lesssim 4 \times 10^{-7} \, ,
\end{equation}
which implies that we remain in a close neighbourhood of the minimum of the scalaron potential.

The adiabaticity constraint \eqref{adcon} reads
\begin{equation}
{\cal P}_\phi \equiv \left\langle \left( \frac{\delta \rho_\phi }{\rho_\phi} \right)^2 \right\rangle = \frac{4 \left\langle \delta \xi_\text{i}^2 \right\rangle}{\xi_\text{i}^2} = \left( \frac{H_\text{inf}}{\pi \xi_\text{i}} \right)^2 \lesssim 10^{-11} \, , 
\end{equation}
together with \eqref{con} implying
\begin{equation}\label{H2}
H_\text{inf} \lesssim 3 \times 10^6\, \left( \frac{g_\text{i}}{100} \right)^{1/8} \left( \frac{\text{eV}}{m} \right)^{1/4}~\text{GeV} \, .
\end{equation}
This gives an upper bound on the reheating temperature: 
\begin{equation}\label{Tr2}
T_r \lesssim 10^{12} \left( \frac{\text{eV}}{m} \right)^{1/8}~\text{GeV} \, .
\end{equation}
In view of the upper bound \eqref{mup} on the mass $m$, the lowest possible bound on the Hubble parameter is $H_\text{inf} \lesssim 10^5\, \text{GeV}$, and that on the reheating temperature is $T_\text{r} \lesssim 2 \times 10^{11}\, \text{GeV}$.

The bounds \eqref{Tr1}, \eqref{Tr2} on the inflationary energy scale, although somewhat low, are not too restrictive.

\section{Summary}

It is remarkable that the scalaron of the generic $f(R)$ gravity can be a dark-matter candidate  \cite{Cembranos:2008gj, Cembranos:2015svp, Shtanov:2021uif}. As in other cases where the role of dark matter is played by a quasiclassical scalar field (e.g., the axion), the present model has the issue of initial conditions.  In the pioneer work \cite{Cembranos:2008gj}, the initial conditions for the scalaron were set somewhere at the stage of the hot evolution of the universe, where $3 H \gtrsim m$.  Their origin and possible primordial spatial inhomogeneities of the scalaron were not analysed. In our paper \cite{Shtanov:2021uif}, we took into account the modification of the scalaron effective potential by its interaction with the environment (notably, the Higgs field) and proposed a scenario in which the scalaron initially resides at the minimum of its effective potential, with the subsequent evolution of the Higgs-field expectation value during the electroweak crossover triggering the excitation of the scalaron. This initial condition has the merit of being special; however, it is hard to envisage its realisation in the early universe. Indeed, the extremely weak interactions of the scalaron, which make it a perfect candidate for dark matter, at the same time preclude its relaxation to the minimum of its effective potential. 

Therefore, in the present paper, we addressed the issue of the initial conditions for the scalaron stemming from inflation. A simple example of the massive inflaton field minimally coupled to gravity in the Jordan frame reveals that the scalaron potential is likely to be strongly modified during inflation, typically becoming non-monotonic. The scalaron is subject to quantum diffusion on super-Hubble spatial scales during inflation, and its value at the end of inflation in the observable patch of the universe is going to be quite random.  Thus, the initial value of the scalaron is unpredictable in this scenario, and one has to rely on the anthropic principle in the model where it constitutes dark matter.

Quantum fluctuations of the scalaron at the final stages of inflation lead to formation of isocurvature primordial perturbations in dark matter. The power spectrum of such perturbations is tightly constrained by current observations. This allowed us to place constraints on the energy scale of inflation compatible with the dark-matter model under consideration.  These constraints turned out to be low but not too restrictive.  Essentially, the Hubble parameter of the last stages of inflation should not exceed $10^5$--$10^7$~GeV [see \eqref{H1}, \eqref{H2}] with the reheating temperature not exceeding $10^{11}$--$10^{12}$~GeV [see \eqref{Tr1}, \eqref{Tr2}]. These results strengthen the case for the viability of the dark-matter model under consideration.

\section*{Acknowledgements}

This research was funded by the National Academy of Sciences of Ukraine under project 0121U109612 and by the Taras Shevchenko National University of Kiev under project 22BF023-01.

\end{document}